\documentclass[pageno]{jpaper}
\pdfoutput=1
\usepackage{cite}
\usepackage{amsmath,amssymb,amsfonts}
\usepackage{algorithmic}
\usepackage{graphicx}
\usepackage{array}
\usepackage{textcomp}
\usepackage{xcolor}
\usepackage{flushend}

\usepackage{algorithmic}
\usepackage[ruled,vlined,linesnumbered]{algorithm2e}

\usepackage{hyperref}
\usepackage{comment}

\usepackage{tikz}
\usepackage{xcolor}
\usepackage{pifont}
\usepackage{tcolorbox}
\usepackage{wrapfig}

\newcommand{\cmark}{\ding{52}}
\newcommand{\xmark}{\ding{56}}%

\newcommand*\circled[1]{\tikz[baseline=(char.base)]{
            \node[shape=circle,fill,inner sep=1pt] (char) {\textcolor{white}{#1}};}}

\newcommand{\hoda}[1]{\textcolor{red}{Hoda: \em #1 }}
\newcommand{\ghadeer}[1]{\textcolor{blue}{Ghadeer: \em #1 }}

\usepackage[normalem]{ulem}

\begin{document}

\title{
  Exploiting Parallel Memory Write Requests for Covert Channel Attacks in Integrated CPU-GPU Systems}

\author{%
  {\rm Ghadeer\ Almusaddar} \\
  galmusa1@binghamton.edu  \\
Binghamton University
\and
{\rm Hoda Naghibijouybari}\\
hnaghibi@binghamton.edu \\
Binghamton University
} 

\date{}
\maketitle

\thispagestyle{empty}

\begin{abstract}
In heterogeneous SoCs, accelerators like integrated GPUs (iGPUs) are integrated on the same chip as CPUs, sharing the memory subsystem. In such systems, the massive memory requests from throughput-oriented accelerators significantly interfere with CPU memory requests. In addition to the large performance impact, this interference provides an attacker with a strong leakage vector for covert attacks across the processors, which is hard to achieve across the cores in a multi-core CPU. In this paper, we demonstrate that parallel memory write requests of the iGPU and more specifically, the management policy of the write buffer in the memory controller (MC) can lead to significantly stalling CPU memory read requests in heterogeneous SoCs.

We characterize the slowdown on the shared read and write buffers in the memory controller and exploit it to build a cross-processor covert channel in Intel-based integrated CPU-GPU systems. We develop two attack variants that achieve a bandwidth of 1.65 kbps and 4.41 kbps and error rates of 0.49\% and 4.32\% respectively. 

\end{abstract}


\section{Introduction}
Microarchitectural covert and side channel attacks are proving to be an increasingly serious threat in traditional CPU-based systems. Such attacks happen when a compromised or untrusted process is co-located with a trusted process on the same hardware. As a result, it exploits the shared hardware resource to leak sensitive information. Many variants of covert and side channel attacks have been explored and studied on a wide range of microarchitectural resources, including but not limited to caches, branch predictors, interconnects and ports, re-order buffer, TLBs and main memory resources ~\cite{percival-05, mehmet-2016, liu-15, gruss-DIMVA-2016, Yarom-2014, branchscope, evtyushkin-16-branch, leaky_buddies,lotr,port_ieeesp,smotherspectre_ccs19, ben_tlb_usenix18, mismanaged_asiaccs21, drama_usenix, lee_dram}. 

Heterogeneous computing systems are proliferating. Such systems combine a number of CPU cores to run the operating system and traditional workloads along with specialized or general-purpose accelerators to perform application-specific computations. While accelerators in heterogeneous systems offer superior performance and power efficiency, they also expose new vulnerabilities and security issues due to their complex architecture and their integration with the rest of the system.

In a heterogeneous System-on-Chip (SoC), accelerators are tightly integrated on the same die as CPU cores and share the on-chip interconnect, last-level cache (LLC) (in some architectures), as well as the main memory and its MC. This tight integration provides high-speed and power-efficient data transfer between the CPU and integrated accelerators. However, it also creates the potential for new attacks that exploit common resources to create interference between these components, leading to cross-component microarchitectural attacks. A recent work~\cite{leaky_buddies} has shown that this resource sharing can lead to microarchitectural covert channels between CPU and iGPU in an Intel-based SoC. Specifically, they build cross-component covert channels through the shared LLC and ring interconnect.

Our work targets the shared MC resources (i.e. read and write buffers) in a heterogeneous SoC (specifically an Intel SoC with iGPU) to build cross-component covert channel attacks. Prior work~\cite{cornell_defense_hpca} proposed a defense scheme against contention-based channels that target shared MC and memory subsystem in a multi-core CPU. To motivate their protection scheme, they showed covert and side channel attacks examples. However, they neither characterize the source of contention nor develop an end-to-end attack. 
Also, there is no indication of the required number of requests to cause the timing difference in both attacks.

In this paper, we study the effect of MC resource characteristics in a modern heterogeneous SoC and investigate the impact of memory write requests from an accelerator on the shared read and write buffers within the MC. We demonstrate that throughput-oriented accelerators with Single-Instruction Multiple-Data (SIMD) architecture which generate massively parallel memory accesses increase the latency of co-running CPU process significantly. We show that such high latency which is experienced by CPU processes can be exploited to construct reliable and high-quality covert channel attacks across the components in a chip.

In comparison with homogeneous multi-core CPUs, heterogeneous systems provide attackers with unique opportunities. For example, massive parallelism in accelerators reduces the time required for filling shared buffers. Such buffers are originally designed with consideration to common characteristics of CPU process memory traffic. Moreover, this new class of microarchitectural attacks (cross-component attacks) relieves the attacker from co-location on the same processor and can bypass existing defense mechanisms that mainly focus on one component of the system. However, new challenges are introduced due to the heterogeneity of components. This includes frequency disparity of CPU and accelerators which makes the covert channel synchronization more challenging, and programming APIs with no (or limited) system support (e.g. lack of timing interface and {\fontfamily{pcr}\selectfont clflush} in OpenCL API). Additionally, exploiting write memory requests due to updated cache lines at the MC level requires special techniques. This is because the CPU process does not know the exact time when these dirty cache lines will reach the shared MC. 

To our knowledge, this is the first attack to target read and write buffers in the MC in heterogeneous SoCs. Furthermore, this is the first work to characterize the high latency of co-running CPU processes due to iGPU memory write requests. This provides important insights into how this threat model manifests in such systems and extends our understanding of the threat to guide future research into defense mechanisms. 

In summary, our paper makes the following main contributions:
\begin{itemize}


\item We identify that parallel memory writes from an integrated accelerator can be a strong leakage vector to build cross-component covert channels through the shared MC. 

\item We characterize the slowdown of memory read requests from co-running CPU process during iGPU kernel memory write requests.

\item We identify the main root of the slowdown is due to the management policy of the write buffer (\textit{drain\_when\_full}) in the shared MC.

\item We develop two covert channel attack variants between CPU and iGPU in Intel-based SoCs by exploiting parallel memory writes and the slowdown caused by the management policy of the write buffer.
\end{itemize}

Our work mainly focuses on developing covert channel attacks. The presence of a covert channel can also forecast the possibility of a side-channel attack to reveal secret information (if the secret is dependent on accelerators' memory write requests).

The paper is organized as follows: Sections~\ref{sec:background}, \ref{sec:threat_model} and \ref{sec:motivation} present background, threat model, and our attack motivation. In Section~\ref{sec:cont_src} we explain the reverse engineering process of the source of the slowdown. Section~\ref{sec:attack_design} explains our attack design, and in Section~\ref{sec:eval} we evaluate our covert channel attack variants. Sections~\ref{sec:discuss} and \ref{sec:mitigations} are discussion and possible mitigations, and Section~\ref{sec:related_work} presents related work. Finally, Section~\ref{sec:conc} concludes our paper.

\section{Background}
\label{sec:background}


We develop our attacks on an Intel-based CPU-GPU system since Intel iGPUs are the most common GPUs with the largest market share~\cite{gpu-market-share}; given that every Intel-based CPU in desktop PCs, workstations, or laptops is equipped with an iGPU. In this section, we introduce Intel's integrated graphics architecture and provide the necessary background for our attacks.

\subsection{Intel Integrated Graphics Architecture}

\begin{figure} [t]
    \centering
    \includegraphics[width=0.35\textwidth, height=0.15\textwidth]{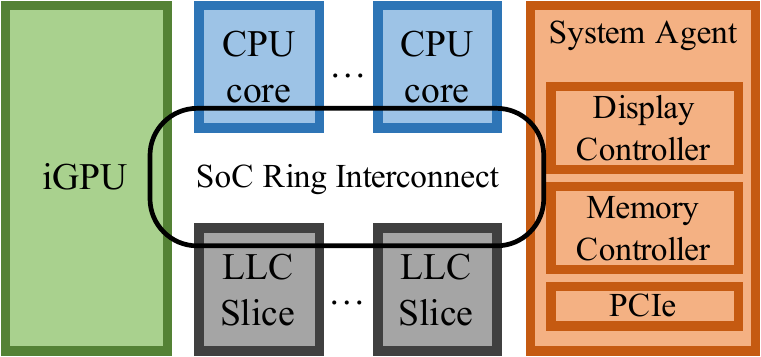}
    \caption{Overall architecture of Intel SoCs.}
    \label{fig:intel_soc}
     \vspace{-2mm}
\end{figure}

\begin{figure} [t]
    \centering
    \includegraphics[width=0.4\textwidth]{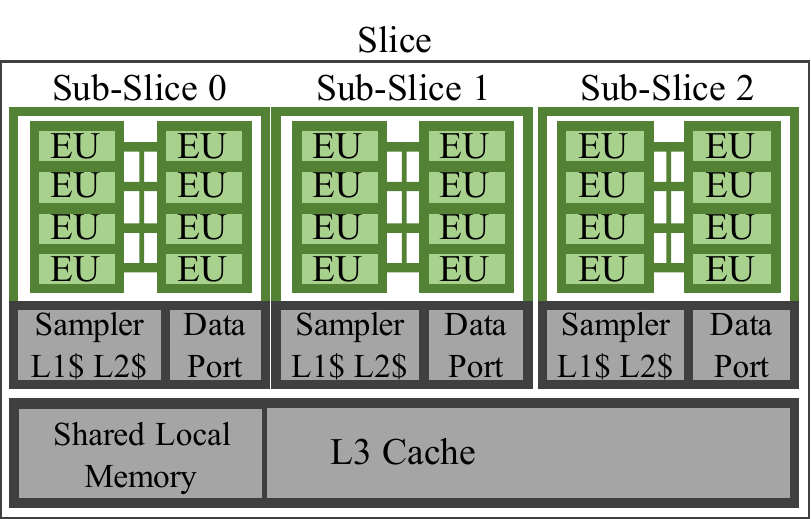}
    \caption{Architectural overview of Intel integrated graphics. 
    }
    \label{fig:igpu-arch}
    \vspace{-2mm}
\end{figure}

In some Intel processors, multiple CPU cores are tightly coupled with GPU on the same chip. Such components in addition to other components such as MC, Last level cache (LLC) slices, and system agent are connected via ring interconnect as we show in Figure \ref{fig:intel_soc}.


Intel iGPU consists of a number of slices. Figure~\ref{fig:igpu-arch} demonstrates the internal architecture of a slice in iGPU. Each slice includes a number of subslices and each subslice contains a number of Execution Units (EUs), each capable of executing seven hardware threads simultaneously. 
Gen9 and Gen9.5 iGPUs contain one slice with three subslices, each with eight EUs (a total of 24 EUs in the slice)~\cite{gen7.5, gen9, gen11}. From the memory hierarchy aspect, each subslice includes read-only L1 and L2 sampler caches that are used only for the graphics stack. EU also includes a data port which is a load and store unit. 
All subslices share an L3 cache which is found to be non-inclusive \cite{leaky_buddies, macsim_ispass}. 
Intel iGPU shares the LLC with CPU cores as well.

\begin{wrapfigure}[9]{r}{0.26\textwidth}
\vspace{-0.5cm}
    \includegraphics[width=0.26\textwidth]{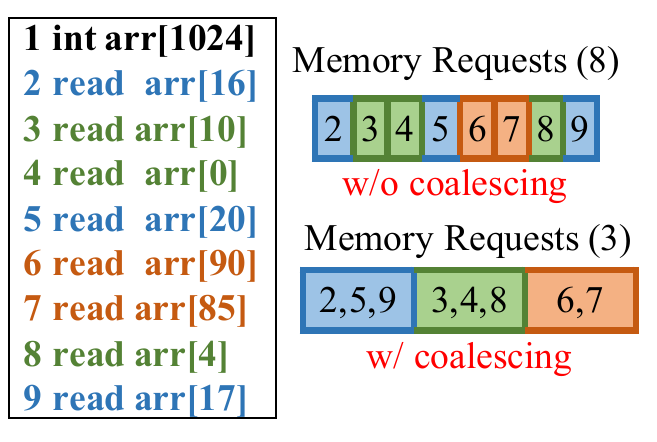}
    \caption{Memory Coalescing.
    }
    \vspace{-2mm}
    \label{fig:mem_coal}
\end{wrapfigure}%

Memory coalescing is one of the approaches used to reduce the memory bandwidth consumption of SIMD
-based workloads. With memory coalescing, memory accesses from threads within the same warp or wavefront (i.e. threads executing the same instruction) which are targeting the same cache line are merged as a single memory request. With such an approach, in addition to improving the performance, it hides memory accesses of different threads within a wavefront due to merging. Figure \ref{fig:mem_coal} explains the process of memory coalescing.

\subsection{Execution Model and Kernel Life Cycle}
We use OpenCL (a cross-platform API) to program iGPU kernels~\cite{opencl}. We next explain about OpenCL execution model and iGPU kernel life cycle.

GPUs operate in SIMD mode, in which a group of work-items (threads) executes the same instruction on multiple data in parallel. Each GPU application consists of some GPU kernels. Each kernel launches a group of threads called work-group in OpenCL terminology. Work-groups are assigned to subslices in a round-robin fashion. Each work-group is broken into sub-work groups (8, 16 or 32 work-items), called wavefront.

The process of executing an iGPU kernel starts from a CPU process. Using OpenCL API, a CPU process enqueues a command in the command queue (CQ) to execute the targeted iGPU kernel. Additionally, buffers to be copied to kernel memory address space are also enqueued in CQ accordingly. At the end of kernel execution, buffers required by the CPU process (i.e. execution result of the iGPU kernel) are enqueued in CQ.

\subsection{LLC Write-backs and Write Buffer Management Policies}
There are multiple approaches that govern the process of writing cache lines in non-blocking caches to main memory due to writes. All dirty cache lines of LLC are added to a write buffer and are eventually written to memory. 
In the write-back scenario, dirty cache lines are grouped and written to memory based on specific events. In this case, a write-back buffer is required which holds dirty cache lines evicted from LLC to be added to MC write buffer.

There are different proposed policies to manage write buffers in memory controller. Such management policies are like \textit{drain-when-full} which empty the write buffer when it is full and stall serving memory reads upon this~\cite{lee_dram, staged_reads}. Another management policy gives priority to reads, such that when the read buffer has memory requests, it stalls serving memory write requests until memory reads are served.

\section{Threat Model}
\label{sec:threat_model}


Our threat model is for a cross-processor covert channel attack in Intel-based SoC (from iGPU to CPU). In a covert channel, two processes (a trojan as a sender and a spy as a receiver) communicate covertly using a shared resource. The spy runs on a CPU core and the trojan process (from another CPU core) launches the kernel through user-level OpenCL API calls to run on iGPU. Both spy and trojan need to execute on the same machine and share the MC. We assume these processes are separate user space processes and do not require any privileged support. Also, there is no shared data between spy and trojan processes.

MC serves read and write memory requests from read and write buffers. Sharing of other architectural and micro-architectural resources other than the MC is not required in our attack. Also, it is not required for CPU cores and iGPU to share the same cache as long as the same MC is serving memory requests.
Our attack targets SoCs with a single MC with dual channels which is the common case with most SoCs in desktop, workstation, and mobile processor chips \cite{gen7.5, gen9, gen11, intel_uhd, arm_npu1, snapdragon_660, qualcomm_660}. All memory requests from the CPU and iGPU are routed through this shared MC from the cache write-back buffer. We exploit the write buffer within the shared MC to develop our covert channel attack variants.

\begin{figure} [t]
    \centering
    \includegraphics[width=0.45\textwidth]{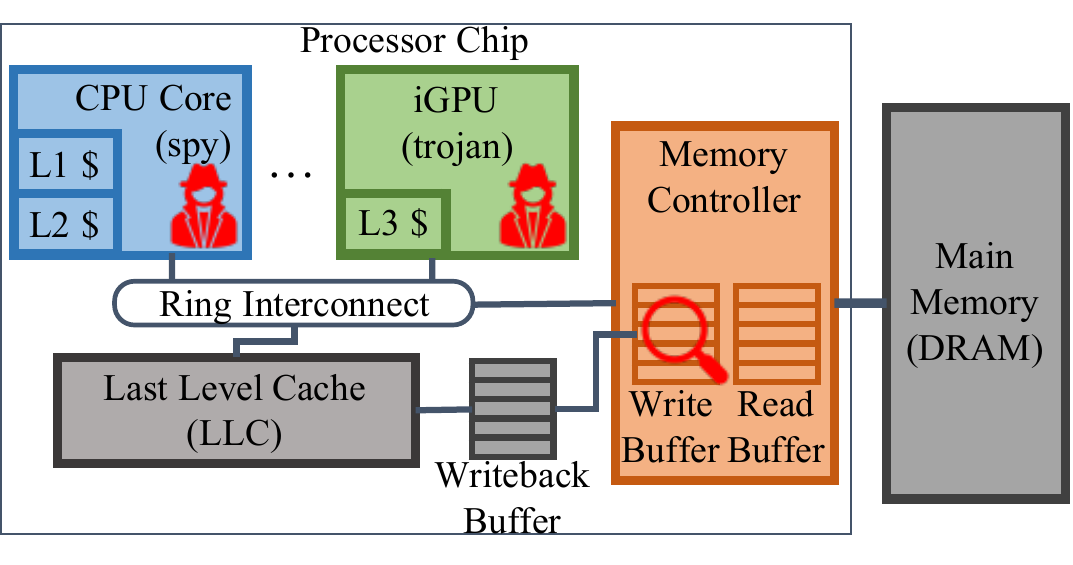}
    \caption{Threat model
    }
    \vspace{-3mm}
    \label{fig:threat-model}
\end{figure}

Our attacks are developed and tested on an unmodified system. Current Intel iGPUs are not capable of running multiple computation kernels from separate contexts concurrently and therefore no noise is expected on the GPU side. Figure \ref{fig:threat-model} demonstrates our covert channel threat model.

\section{Motivation and Challenges}
\label{sec:motivation}
In this section, we motivate our attack which targets the write buffer in the MC to leak secret information. Also, we strengthen the point of using iGPU and write buffer compared to other possible shared resources.

\subsection{Memory Controller's Write Buffer}
Our attack exploits the write buffer in the MC. MC is shared between CPU cores and iGPU. There are many incentives for attackers to target MC resources. Buffers in the MC are shared between all CPU cores and iGPU in some heterogeneous SoCs. Consequently, they are not affected by existing cache-based defense approaches such as partitioning or randomization~\cite{pl_cache, scatter_cache}. Buffers within MC are usually smaller in size than caches, they do not have sophisticated replacement policies since they act as queues, and do not depend on address mapping and index hashing (e.g. across LLC slices in the case of Intel processors). As a result, overflowing or saturating these buffers is straightforward and depends on the frequency of memory requests. Furthermore, memory requests in buffers/queues within the MC are generally serviced using common scheduling policies. All these factors make targeting buffers in MC tempting for attackers to leak secret information.

There are multiple challenges that make targeting the MC's write buffer challenging for attackers. 
At the level of the MC, the write buffer status does not directly indicate the current execution status of the program. This is because entries in the write buffer could be due to an earlier execution state. Furthermore, the management policy of the write buffer plays a role in serving memory requests such as in the case of a full buffer or when there is standing read requests. Triggering write buffer related events such that secret information is leaked is dependent on write buffer size, its management policy, and the frequency of memory write requests as we will show later. Also, we will show that creating sufficient write requests to drain the write buffer in a secret-dependent way can not be easily achieved using memory traffic from a CPU core. This motivates our work in using iGPU memory traffic in modern heterogeneous systems.
\begin{tcolorbox}[arc=5mm, outer arc=1mm, width=\linewidth,halign=flush center, left=1mm, right=1mm]
\textit{
Write buffer status affects memory read requests.
}
\end{tcolorbox}

    

\subsection{CPU-CPU vs. CPU-iGPU Slowdown}

Our attack targets MC resources, and mainly the write buffer. To our knowledge, this is the first work to target the write buffer using iGPU memory write traffic. In this subsection, we motivate using iGPU memory traffic compared to CPU memory traffic. Later, we compare our attack and threat model with existing work that targeted MC resources, DRAM or LLC, and ring-interconnect.   

\subsubsection{CPU-CPU Slowdown}
To explore CPU-CPU possible slowdown at the MC level, we executed two processes (A) and (B) on different cores. Process(A) is iteratively reading from a buffer. We ensure each buffer access is a memory access by flushing the targeted cache line beforehand using {\fontfamily{pcr}\selectfont clflush} instruction. Process (B) on a different CPU core accesses another buffer. In process (B), we tried both memory reads and writes. For reads and writes, we access different cache lines. In the case of reads, we flush the buffer's cache line before accessing it. We measure the access time on process (A) using {\fontfamily{pcr}\selectfont rdtscp} instruction. 

Figure~\ref{fig:slowdown_a} shows the slowdown observed by CPU process (A) due to memory accesses of CPU process (B) for different numbers of memory requests in the case of reads and writes. The latency values are normalized to the latency of memory accesses when there is no contention (baseline). Slowdown due to the CPU process's memory accesses is observable only at a large number of memory write requests. Memory reads are not causing a notable slowdown similar to the case observed with memory writes.

\subsubsection{CPU-iGPU Slowdown}

We investigated CPU process slowdown which can be caused by iGPU memory traffic. We executed two CPU processes: Process (A) and Process (B). Similar to the CPU-CPU slowdown experiment, process (A) is continuously doing memory read requests. To make sure all access are served from main memory, we used {\fontfamily{pcr}\selectfont clflush} instruction. CPU process's memory access latency is measured using {\fontfamily{pcr}\selectfont rdtscp} instruction. Process (B) is launching iGPU kernel and the possible slowdown is tested due to kernel memory requests from both reads and writes with a different number of memory requests. 

The maximum size of the local work-group in iGPU kernel is 256 work-items (i.e. maximum local work-group size supported by OpenCL API), and the size of the global work-group is set according to the desired number of memory requests. For example, if the targeted number of memory requests is less than 256 requests, then size of global work-group is set to be the same as the number of memory requests, otherwise, the global work-group size is set to 256 work-items (the maximum size). As a result, the number of work-groups is one. This is done to avoid the overhead of synchronizing between different work-groups. Memory read and write requests are performed from/to different cache lines to avoid the effect of memory coalescing and guarantee the required number of memory requests for creating contention.

\begin{figure} [t]
    \subfloat[\label{fig:slowdown_a}]{\includegraphics[width=0.525\columnwidth]{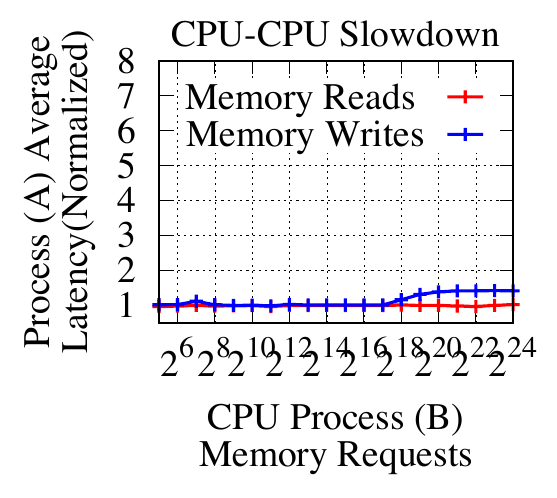}}%
    \subfloat[\label{fig:slowdown_b}]{\includegraphics[width=0.525\columnwidth]{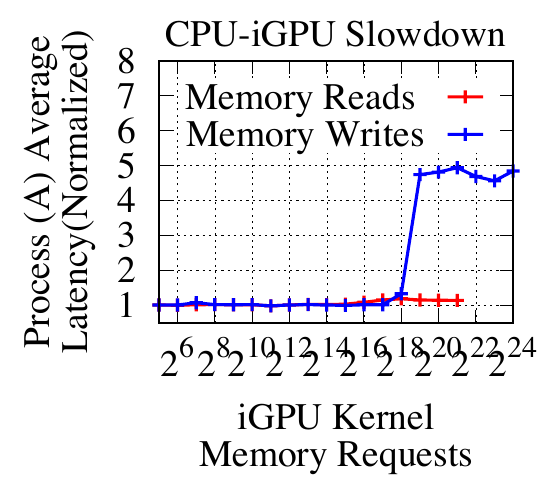}}
    \caption{CPU process (A) normalized latency during execution of (a) CPU process (B) and (b) iGPU kernel for both memory read and write requests.}
    \label{fig:slowdown}
\end{figure}

As it can be noticed from Figure~\ref{fig:slowdown_b}, CPU process (A) slowdown is clearly observable in the case of iGPU kernel's memory write requests. It is barely notable in the case of the kernel's memory read requests. We elaborate on the reason behind such a slowdown through reverse-engineering in the next section. The slowdown caused by the iGPU kernel's memory writes is larger ($\times5$) than the slowdown caused by the CPU process's memory accesses. Also, such a slowdown is caused by using less number of memory write requests compared to the CPU process. This motivates the use of iGPU kernel memory traffic for leaking secret information rather than the CPU core's memory requests. 


\begin{figure} [t]
    \hspace{-5mm}
    \subfloat[\label{fig:llc_cont_hits}]{\includegraphics[width=0.55\columnwidth, height=0.4\columnwidth]{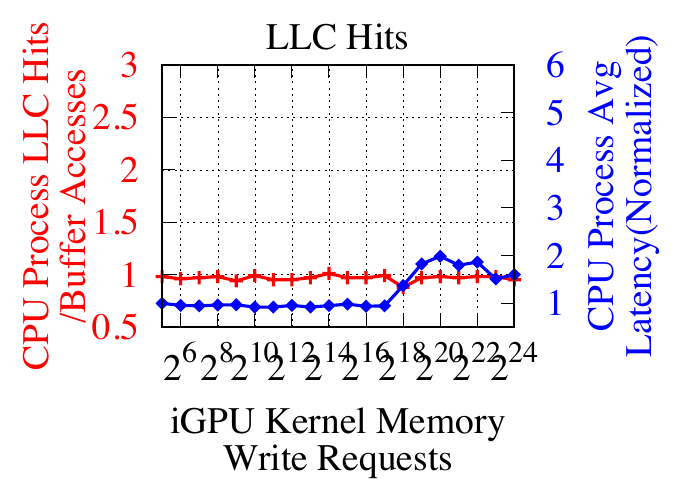}}%
    \subfloat[\label{fig:llc_cont_misses}]{\includegraphics[width=0.55\columnwidth, height=0.4\columnwidth]{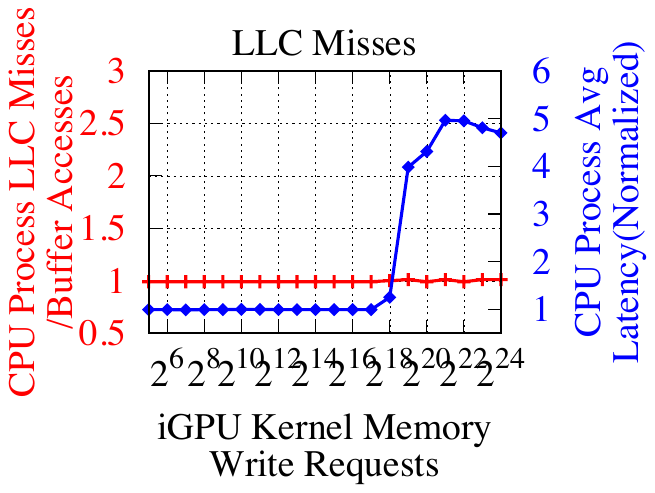}}   
    \caption{Co-running CPU process normalized latency and LLC access status during iGPU kernel memory write requests. CPU process average latency is normalized to the latency when iGPU kernel is not issuing memory write requests. (a) CPU process accesses are LLC hits and (b) CPU process accesses are LLC misses}
    \label{fig:llc_cont}
    \vspace{-5mm}
\end{figure}

\begin{tcolorbox}[arc=5mm, outer arc=1mm, width=\linewidth,halign=flush center, left=1mm, right=1mm]
\textit{
Latency of memory reads from a co-running CPU process during iGPU kernel writes is ($\times5$) larger than the latency during another CPU process writes ($\times0.5$).
}
\end{tcolorbox}

\subsection{Comparison with Related Attacks}

\begin{table*}[t]
    \centering
    \begin{tabular}{||p{2.6cm}|p{2cm}|p{9cm}|p{2cm}||}
    \hline
     {}&\centering{End-to-End Attack}& \centering{Attack Target}& {Attack Parties}\\ 
   
         \hline
        Y. Wang et al.~\cite{cornell_defense_hpca} & \centering\xmark & Memory Controller (No reverse engineering) & CPU-CPU\\ 
         &  & *Main cause of contention/slow-down is not reverse engineered &  \\
        \hline
        DRAMA~\cite{drama_usenix} & \centering\cmark & DRAM (Bank Contention) & CPU-CPU \\ 
        \hline
        Leaky Buddies~\cite{leaky_buddies} & \centering\cmark & LLC \& Ring Interconnect & iGPU-CPU \\ 
        \hline  
        \cellcolor{red!30} Our Attack  & \centering\cellcolor{red!30}\cmark & \cellcolor{red!30} Memory Controller (Read and Write Buffers) & \cellcolor{red!30} iGPU-CPU \\
        \hline
    \end{tabular}
    \caption{Status of our attack compared to related attacks. 
    }
    \label{tab:related_work}
\end{table*}

We further motivate our proposed attack by comparing it to existing attacks that target MC resources and/or integrated accelerators in heterogeneous SoCs.
Our attack is the first to characterize the slowdown (through reverse engineering) and develop end-to-end covert attacks on MC resources  mainly the write buffer.

Wang et al.~\cite{cornell_defense_hpca} proposed a defense approach targeting timing attacks on the MC. To motivate their protection scheme, they briefly discussed the possibility of a covert and a side channel attack without characterizing and confirming the source of contention. For the side channel attack, they targeted RSA decryption algorithm and caused a cache miss when the modulo operation is executed. The modulo is executed when the bit in the private key is one '1'. They showed that the execution time of the attacker increases when the number of bit one in the private key increases. They also showed the possibility of a covert channel attack. In this attack, to send bit one, the adversary issues memory requests, while nothing is issued when sending bit zero.

Some other works studied covert and side channels on the other resources in the memory subsystem rather than the MC. Pessl et al. proposed DRAMA~\cite{drama_usenix}, covert and side channel attacks by utilizing DRAM bank contention. Their attack requires attack parties to agree on set of a rank, a channel, a bank group, and a bank to leak secret information. Our attack does not require this. Dutta et al. ~\cite{leaky_buddies} developed covert channels across CPU and iGPU in Intel-based SoCs, but targeted LLC and ring interconnect which are shared between iGPU and CPU cores. Their attack requires finding precise eviction and polluting sets on LLC to ensure a successful attack.

Table~\ref{tab:related_work} compares our proposed attack against existing attacks that target MC and DRAM resources and/or developed in CPU-iGPU SoCs.

\section{Reverse Engineering: Source of CPU Process's Read Latency Overhead}
\label{sec:cont_src}

CPU cores and Intel iGPU share microarchitectural components other than MC, such as Ring interconnect, LLC, and DRAM resources. Consequently, it is possible for such resources to be the reason for the slowdown observed by the process running on the CPU core during iGPU kernel memory writes. In this section, we investigate the source of slowdown due to iGPU kernel memory writes and confirm that it is due to the management policy of the write buffer in the shared MC. 

In all of our experiments in this section, the CPU process is reading from a buffer of size 128KB except for the experiment for LLC hits (in Figure~\ref{fig:llc_cont_hits}) where the buffer size is 512KB (double the size of L2 cache). We choose a 128KB buffer size to simplify the process of reverse engineering by reducing the set of read addresses from the CPU process (2048 accesses). While a 512KB buffer is used to ensure that most of the accesses from the CPU process are LLC hits.
\subsection{Ring Interconnect and Last Level Cache (LLC)}%
In our targeted system, LLC and ring interconnect are shared between CPU and iGPU. These two components could be the cause of the slowdown the CPU process is observing. We show that neither LLC nor ring interconnect is the source of contention observed in Figure~\ref{fig:llc_cont}.

We explored the slowdown due to iGPU kernel memory write traffic when CPU process accesses are LLC hits vs. LLC misses. Figure~\ref{fig:llc_cont_hits} shows the normalized latency of CPU process during iGPU kernel writes when buffer accesses are LLC hits. Note that in Figure~\ref{fig:llc_cont_hits} the baseline latency is CPU process average latency when iGPU kernel is not issuing any memory write requests. 
Figure~\ref{fig:llc_cont_misses} shows the normalized latency of the CPU process during iGPU kernel writes when buffer accesses are LLC misses. The baseline for Figure~\ref{fig:llc_cont_misses} is CPU process average latency when the iGPU kernel is not issuing memory write requests. 

It can be noted from Figure~\ref{fig:llc_cont}, that the CPU process is suffering a higher level of slowdown in case of LLC misses ($\times5$ baseline LLC miss latency) compared to LLC hits ($\times2$ baseline LLC hit latency). This proves that neither LLC nor ring-interconnect is the source of slowdown. The higher latency starts to appear when the number of iGPU kernel memory requests is $2^{19}$ write requests, and iGPU kernel buffer size is 32MB.

\begin{tcolorbox}[arc=5mm, outer arc=1mm, width=\linewidth,halign=flush center, left=1mm, right=1mm]

\textit{\textbf{LLC and Ring Interconnect are not the cause of CPU process's slowdown.} During iGPU kernel execution, the normalized latency of the co-running CPU process when its accesses are LLC hits is much lower than the normalized latency when its accesses are LLC misses. 
}
\end{tcolorbox}

\subsubsection{Writeback Buffer of LLC}
Writeback buffer of LLC stores dirty cache lines to be up- dated in main memory. Due to the high rate of memory write requests by the iGPU kernel, writeback buffer of LLC will rapidly get filled. We showed in Figure~\ref{fig:llc_cont_hits} that the latency of the co-running CPU process when buffer accesses are LLC hits is $\times2$ baseline LLC hit latency. This is about $\times12$-$\times15$ smaller than the latency when CPU process buffer accesses are LLC misses. This indicates that LLC does not get blocked from serving CPU process read requests even though the number of write memory requests is the same in Figure~\ref{fig:llc_cont_hits} and Figure~\ref{fig:llc_cont_misses}. Also, the LLC writeback buffer is not on the critical path of process's read requests whether they are hit or miss in LLC as long as they are not dependent on write requests.

\begin{tcolorbox}[arc=5mm, outer arc=1mm, width=\linewidth, halign=flush center, left=1mm, right=1mm]
\textit{\textbf{Writeback buffer of LLC is not the source of CPU process's slowdown}. This is because the Writeback buffer is not on the critical path of memory read requests which are not dependent on writes.}
\end{tcolorbox}

\begin{table}[t]
\centering
    \begin{tabular}{||m {0.33\columnwidth}|m{0.56\columnwidth}||}
        \hline
        Memory Controller & Dual channel \\
        \hline
        DRAM & DDR4 MR[ABC]4U320GJJM16G @ 2600MT/s\\        
        \hline
        Memory Capacity & 32GB (2-16GB DIMMs) \\
        \hline
        Ranks & Single Rank \\
        \hline
        Number of Bank groups and Banks & 4 bank groups, 4 banks/bank group \\
        \hline
        Channel Addressing & $b_8\oplus b_9\oplus b_{12}\oplus b_{13}\oplus b_{15} \oplus b_{16}$\\
        \hline
        Bank Group Addressing & BG0: $b_7\oplus b_{14}$ \newline BG1: $b_{15}\oplus b_{18}$\\
        \hline
        Bank Addressing & BA0: $b_{16}\oplus b_{19}$ \newline BA1: $b_{17}\oplus b_{20}$ \\
        \hline
    \end{tabular}
\caption{Targeted DRAM details and reversed engineered channel, bank group and bank addressing.}
\label{tab:dram_re}
\vspace{-2mm}
\end{table}

\subsection{Memory Controller and DRAM Resources}

\begin{figure*} [t]
    \subfloat[\label{fig:latency_a}]{\includegraphics[width=0.5\columnwidth]{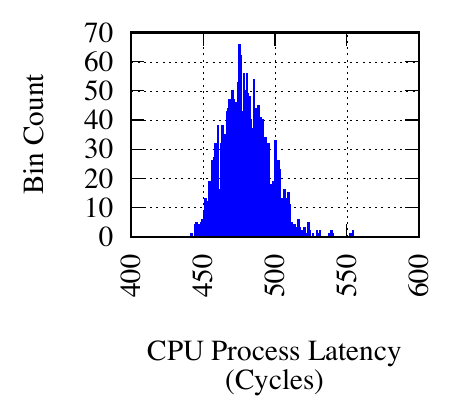}}
    \subfloat[\label{fig:channel_a}]{\includegraphics[width=0.5\columnwidth,height=0.46\columnwidth]{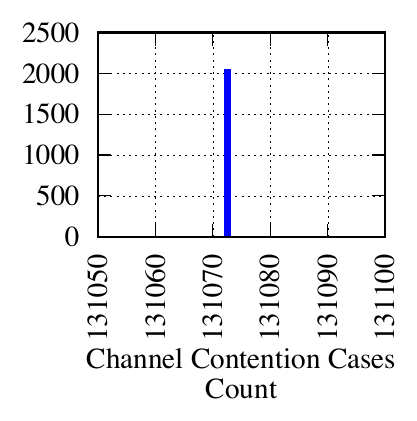}}
    \subfloat[\label{fig:bankgroup_a}]{\includegraphics[width=0.5\columnwidth,height=0.46\columnwidth]{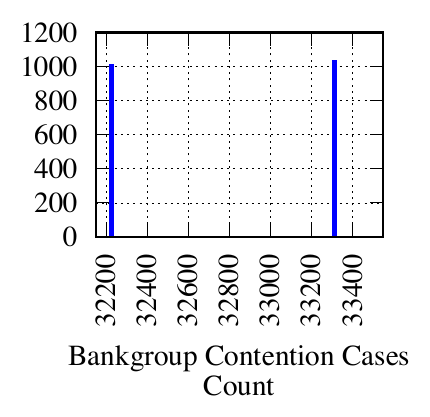}}
    \subfloat[\label{fig:bank_a}]{\includegraphics[width=0.5\columnwidth,height=0.46\columnwidth]{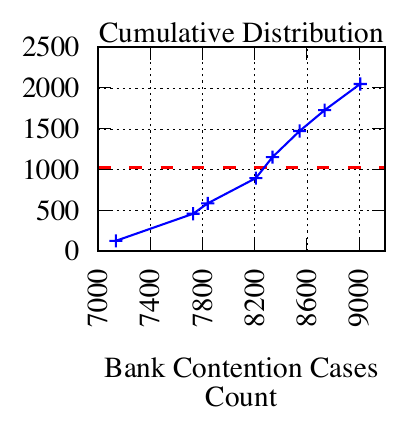}}
    \caption{(a) CPU process latency distribution, (b) Channel contention cases distribution, (c) Bank group contention cases distribution and (d) Bank contention cases cumulative distribution. iGPU kernel is issuing $2^{18}$ memory write requests by accessing \textbf{\underline{all cache lines of 16MB buffer}}, and CPU process is issuing 2048 memory read requests.}
    \label{fig:dram_cont_a}
    \subfloat[\label{fig:latency_b}]{\includegraphics[width=0.5\columnwidth,height=0.46\columnwidth]{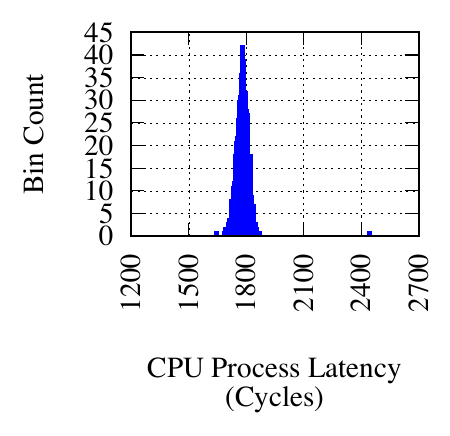}}
    \subfloat[\label{fig:channel_b}]{\includegraphics[width=0.5\columnwidth,height=0.46\columnwidth]{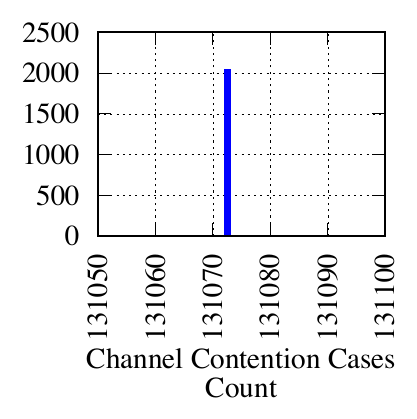}}
    \subfloat[\label{fig:bankgroup_b}]{\includegraphics[width=0.5\columnwidth,height=0.46\columnwidth]{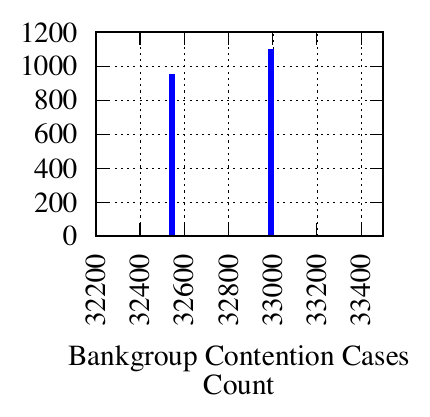}}
    \subfloat[\label{fig:bank_b}]{\includegraphics[width=0.5\columnwidth,height=0.46\columnwidth]{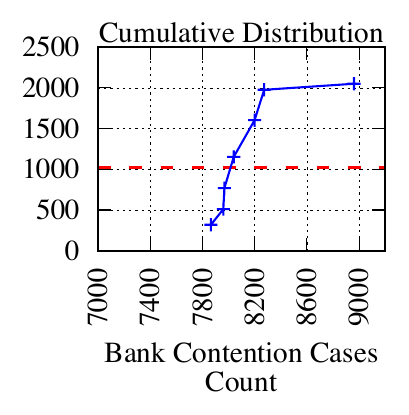}}    
    \caption{(a) CPU process latency distribution, (b) Channel contention cases distribution, (c) Bank group contention cases distribution and (d) Bank contention cases cumulative distribution. iGPU kernel is issuing $2^{18}$ memory write requests by accessing a \textbf{\underline {32MB buffer at a stride of one cache line}}, and CPU process is issuing 2048 memory read requests.
    }
    \label{fig:dram_cont_b}
\end{figure*}

CPU cores and iGPU also share MC and main memory (DRAM) resources. We investigate whether shared resources such as channels, bank groups, or banks contribute to the high latency level observed by the co-running CPU process during iGPU kernel memory write requests.

We reversed engineered bits in the physical address which indicate the rank, channel, bank group, and bank to access based on DRAMA paper ~\cite{drama_usenix}. Table~\ref{tab:dram_re} shows reverse engineering results of the channel, bank group, and bank addressing used to determine which DRAM resource is used based on the physical address. The table also shows our targeted DRAM details.

To study the impact of the channel, bank group, or bank contention on the observed high latency, we launched two experiments (A) and (B). In both experiments, the CPU process and iGPU kernel are doing the same number of memory reads and writes; memory reads in the case of the CPU process and memory writes in the case of the iGPU kernel. The difference is that we allocated a larger buffer in experiment (B) which is accessed at a stride of one cache line.

Figure~\ref{fig:dram_cont_a} and Figure~\ref{fig:dram_cont_b} depict CPU process read latency distribution and channel, bank group, and bank contention cases distribution. A contention case happens when CPU process read address uses the same channel, same bank group or same bank as iGPU kernel memory write addresses based on physical addresses. We infer if a channel, bank group, or bank contention case had happened based on reverse-engineered addressing using physical address bits in Table \ref{tab:dram_re}.

In these experiments, the total number of read requests by the CPU process is 2048 requests and the total number of write requests for the iGPU kernel is $2^{18}$ memory requests. In Figures~\ref{fig:channel_a}, ~\ref{fig:channel_b}, ~\ref{fig:bankgroup_a}, ~\ref{fig:bankgroup_b}, ~\ref{fig:bank_a} and ~\ref{fig:bank_b}, the x-axis indicates the number of contention cases observed by each CPU process read access due to iGPU kernel memory writes. The y-axis indicates the count of these contention cases. Note that the total of contention cases count (y-axis) is 2048 which is equal to CPU process memory accesses.


Figure~\ref{fig:channel_a} and Figure~\ref{fig:channel_b} show the distribution of channel contention cases observed by the CPU process due to iGPU kernel memory writes. All of the CPU process memory reads observed the same number of channel contention cases equal to half of iGPU kernel write requests in both experiments ($2^{17}$). This is because based on our reverse engineering results, we found that about half of the iGPU kernel buffer was allocated on the first channel and the other half to the second channel. The same scenario is for the CPU process buffer. We conclude that channel contention is not the reason behind the high latency observed by the CPU process since channel contention level is the same in experiments (A) and (B).

Figure~\ref{fig:bankgroup_a} and Figure~\ref{fig:bankgroup_b} depict the distribution of bank group contention cases observed by the CPU process due to iGPU kernel memory write requests. In Figure~\ref{fig:bankgroup_a}, about 1000 CPU process memory accesses observed 32200 contention cases, while the rest observed 33300 contention cases. The case for Figure~\ref{fig:bankgroup_b} is close; about 950 memory accesses suffered 32550 contention cases the rest suffered 33000 contention cases. Total bank group contention cases in both experiments are close and do not explain the large difference in latency distribution between experiments (A) and (B).

Furthermore, we investigate the difference in bank contention between experiments (A) and (B) as we show in Figure~\ref{fig:bank_a} and Figure~\ref{fig:bank_b}. In Figure~\ref{fig:bank_a}, about 50\% of CPU process accesses resulted in 8200 bank contention cases or lower. Most CPU process accesses suffered contention cases between 7800 and 8910 cases. While for the second experiment in Figure~\ref{fig:bank_b}, 50\% of CPU process accesses resulted in about 8000 bank contention cases or lower. Also, most CPU process accesses in this experiment suffered contention cases between 8000 and 8370 cases. The range of contention cases in experiment (B) is smaller than in experiment (A). From these results, we can conclude that bank contention is not the reason behind the huge latency difference between experiments (A) and (B) shown in Figure~\ref{fig:latency_a} and Figure~\ref{fig:latency_b}. 

\begin{tcolorbox}[arc=5mm, outer arc=1mm, width=\linewidth, halign=flush center, left=1mm, right=1mm]
\textit{Channel, Bank group, and Bank contention are not the cause of the high latency of co-running CPU process during iGPU kernel memory writes.}
\end{tcolorbox}

\subsection{Write and Read Buffers in Memory Controller}
Other resources which could increase the latency of memory reads performed by the CPU process are read and write buffers in the MC. The high latency observed by the co-running CPU process happens only in the case of iGPU kernel memory writes not memory reads. Also, we noticed that the slowdown experienced by the CPU process happens once during iGPU kernel kernel buffer accesses when sending single bit '1' and not periodically as we show in Figure~\ref{fig:latency_per_access}. The CPU process continue to suffer high memory access latency during iGPU kernel execution.%

Considering these circumstances, it is possible for the management policy of the write buffer in the MC to be the cause of the CPU process's higher read latency. A common management policy used with write buffer in MC is \textit{drain\_when\_full}. In this management policy, when the write buffer gets full, memory read requests are stalled until the write buffer is drained. The purpose of such a management policy is to avoid DRAM latency due to \textit{write after read} and \textit{read after write}. In fact, there is a lot of research addressing this performance issue which occurs when stalling memory read requests to serve write requests in multi-core CPU environment~\cite{staged_reads, lee_dram}.%

Due to continuous parallel write requests from iGPU, the probability of getting the write buffer full is high. Also, when the write buffer is being drained, there will be a number of standing write requests waiting to be added to the write buffer to be served. This explains the high latency observed by the CPU process when reading from the main memory.
\begin{tcolorbox}[arc=5mm, outer arc=1mm, width=\linewidth, halign=flush center, left=1mm, right=1mm]
\textit{Due to the write buffer management policy (\textit{drain\_when\_full}), memory read requests in the read buffer need to be stalled to serve memory write requests.}
\end{tcolorbox}

\section{Covert Channel Attack Design}
\label{sec:attack_design}

We concluded that the management policy of the write buffer is the cause of the high latency observed by the co-running CPU process. In this section, we explore different ways to exploit this phenomenon to leak secret information using covert channel attacks. 

\begin{figure} [h]
    \centering
    \includegraphics[width=0.4\textwidth]{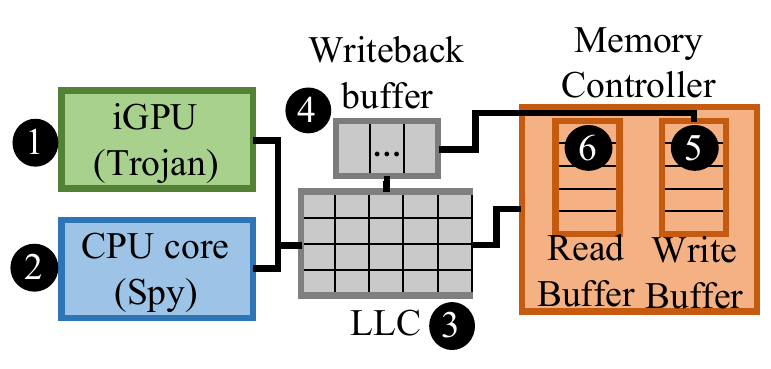}
    \caption{Covert Channel Attack Design.}
    \label{fig:attack_design}
    \vspace{-4mm}
\end{figure}

The attack first starts with handshaking between trojan (CPU-side) and spy. This can be done using traditional techniques (i.e. flush+reload, prime+probe) or by using the slowdown which can be caused using iGPU kernel memory write requests. \circled{1} Trojan CPU process indicates the start of the communication to leak secret information and launches a kernel on the iGPU to issue parallel memory write requests.
\circled{2} Once the spy CPU process successfully detects the bit sequence as part of the handshaking process from trojan (CPU-side), it starts a continuous stream of memory read requests. We found that allocating a buffer with a size larger than the LLC size and accessing it at a stride of 64 cache lines is sufficient to ensure memory accesses during iGPU kernel execution without the need to use clflush instruction. This ensures continuous monitoring of the slowdown caused by the iGPU kernel. \circled{3} Cache lines in LLC will be updated based on iGPU kernel writes. 

\circled{4} When dirty cache lines are evicted from LLC, they are pushed into the write back buffer of LLC. \circled {5} Eventually, dirty cache lines in the write back buffer will be inserted to write buffer in the MC to be updated in main memory. We can control filling the write back buffer and write buffer in the MC by exploiting iGPU parallelism and writing to multiple cache lines in a short time. \circled{6} When the write buffer gets full, it has to be drained due to \textit{drain\_when\_full} management policy. During draining the write buffer, any spy read requests in the read buffer will be stalled until the write buffer is drained.

To send bit "1", trojan has to do a lot of writes to different cache lines to fill and drain write buffer frequently resulting in stalling spy read requests. Writes have to be done to different cache lines to avoid the coalescing effect in iGPU. To send bit "0", trojan exploits coalescing effect, by doing the same number of writes, but to the same cache line to avoid filling the write buffer. This way, the trojan can leak secret information to the spy.

We propose two different approaches to establish a covert channel which exploit the effect of write buffer management policy. The first is MC channel oblivious attack while the second attack targets one MC channel.

\subsection{Attack Variant 1: MC Channel Oblivious Attack}
\label{subsec:var1_design}


\begin{figure}[t]
    \centering
    \includegraphics[width=0.85\columnwidth, height=0.3\columnwidth]{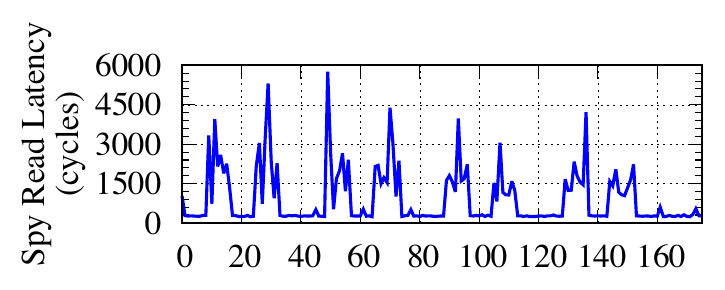}
    \caption{Spy process read latency during iGPU kernel execution to send secret bits. High latency represents bit '1' while low latency represents bit '0'.}
    \label{fig:latency_per_access}
\end{figure}

\begin{figure}[t]
\vspace{-2mm}
    \centering
    \includegraphics[width=\columnwidth]{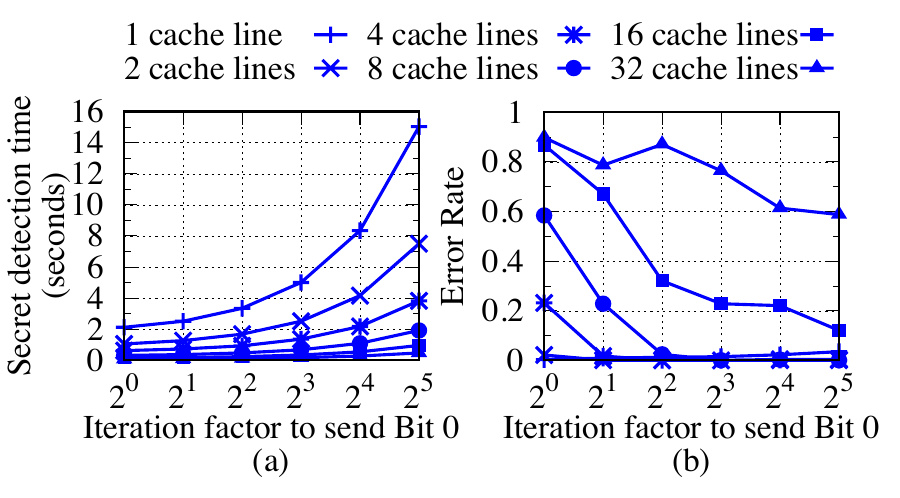}
    \vspace{-0.5cm}
    \caption{(a) Secret detection time and (b) error rate for different bit '0' iterations factors to send bit and at different strides (cache lines) used to access iGPU kernel buffer (32MB).}
    \label{fig:zero_iter_factor}
    \vspace{-2mm}
\end{figure}

\begin{figure}[t]
\hspace{-5mm}
\subfloat[\label{fig:local_threads}]{\includegraphics[width=0.525\columnwidth]{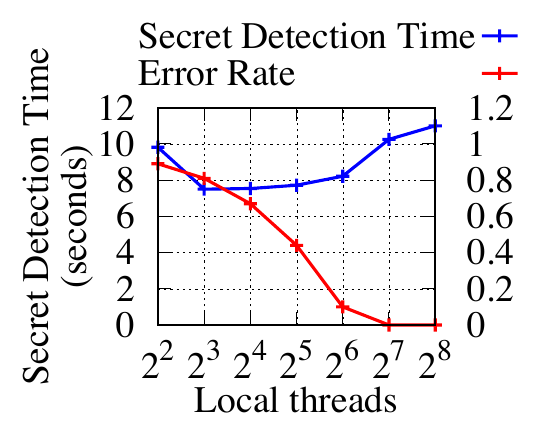}}
\subfloat[\label{fig:global_threads}]{\includegraphics[width=0.55\columnwidth]{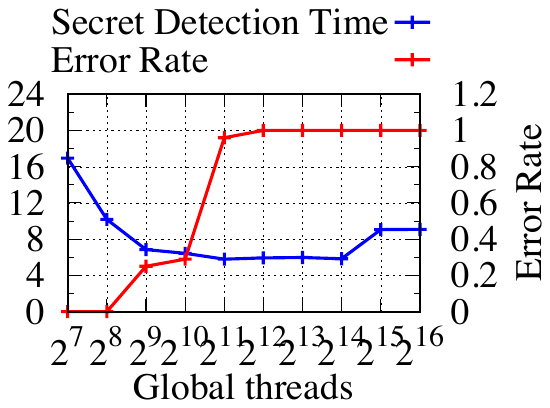}}
\caption{(a) Secret detection time and (b) error rate at different local and global threads. Local threads=size of local work-group. Number of local work-groups=global threads/size of local work-group}
\label{fig:glob_loc_threads}
\vspace{-2mm}
\end{figure}

The first attack variant we are proposing does not require spy and trojan to agree on a channel to access beforehand. Trojan (iGPU kernel) will write to different cache lines to avoid memory request coalescing. From Figure~\ref{fig:slowdown_b}, we can notice the slowdown abruptly appears when the number of write requests is $2^{19}$ memory requests when all cache lines are accessed. Consequently, the allocated buffer size should be 32MB since writes have to be done to different cache lines to avoid the coalescing effect. A buffer size of 32MB is double the size of LLC in our targeted system.

Accessing different cache lines from the allocated 32MB buffer will increase the probability of dirty cache line evictions. Considering the parallelism of iGPU, this will speed up filling the write buffer in the memory controller to stall spy (CPU process) read requests. The drawback of this approach is that it requires iGPU kernel to access a large number of cache lines to evict dirty cache lines. This will lower the bit rate as we will show later. 

To construct an efficient covert channel attack in terms of bit and error rates, we need to consider multiple attack parameters. Such parameters are the number of write requests, the number of local and global threads, and the iteration factor to send bit '0'. We investigated the role of these parameters as we show in Figure \ref{fig:zero_iter_factor} and Figure \ref{fig:glob_loc_threads}.

We mentioned that for the iGPU kernel to send bit '0', it does the same number of writes as in sending bit '1'. We noticed that based on the latency level observed by the spy process when sending bit '1', an iteration factor of one is not enough to send bit '0' after sending bit '1'. We investigated both secret detection time and error rate for different bit '0' iteration factors and different number of write requests as we depict in Figure \ref{fig:zero_iter_factor}.

In Figure \ref{fig:zero_iter_factor}, local and global threads number is 256 and the size of secret is 1024 bits. Increasing the kernel buffer access stride, will decrease the number of write requests and thus decrease spy execution time. However, using larger strides to access iGPU kernel buffer leads to higher error rates since the number of evicted cache lines will be lower. Increasing bit '0' iteration factor shows to decrease the error rate because it makes bit '0' detectable. For this attack and from Figure \ref{fig:zero_iter_factor} we can observe that accessing iGPU kernel at a stride of eight cache lines and with bit '0' iteration factor of eight results in low error rate (~0.1\%) and small execution time.

Furthermore, we investigated the role of number of local and global threads in attack's performance. Local work group size is equal to local threads and the total number of local work-groups is equal to the number of global threads divided by local work-group size. In Figure \ref{fig:local_threads}, the number of global threads is equal to 256 threads. We notice that increasing local threads decrease execution time for initial cases and starts increasing when the number of local threads is 64. In our targeted iGPU, the size of a wavefront (sub-work group size) can be 8,16, or 32. As we increase the number of local threads, the error rate decreases and this is because, with more local threads, more memory write requests are generated. Based on this experiment, we use a local work-group size of 128 threads because it achieves a low error rate and acceptable execution time compared to the case of 64 local threads which achieves ~10\% error rate.

We also explored the role of increasing the number of global threads. The execution time initially decreases until the number of global threads reaches 2048. Error rate becomes ~100\% for global threads equal to or more than 2048 threads. When global threads are 128 or 256 threads, error rate is smaller than 0.2\%. Although 512 threads achieve execution time lower than 128 and 256 threads, it has a higher error rate (~22\%). Having a larger number of global threads, means lower number of writes per thread and this will lower latency overhead caused by iGPU kernel since all of these threads can't be scheduled at the same time. Also, there is the overhead of creating and scheduling these threads.

Based on these experiments, we conclude the parameters required to achieve an efficient covert channel attack which is oblivious of accessed MC channel. iGPU kernel has to write to 32MB buffer at a stride of eight cache lines. Additionally, To make bit zero detectable by the spy process, its iteration factor must be eight considering iGPU kernel buffer size and stride. Also, the local work-group size should be 128 threads and the number of global threads should be 256. As a result, two local work-groups will be created by the kernel for this attack.

\subsection{Attack Variant 2: Targeting Single MC Channel}
\label{subsec:attack_var2}
In our targeted system, CPU core and iGPU share dual channel MC. Each channel has its own read and write buffers. From reverse engineering of MC channel bit addressing, we noticed that the allocated buffer is distributed between these two channels. Considering this, it is possible for spy (CPU process) and trojan(iGPU kernel) to pre-agree on a channel such that single pair of read and write buffers are targeted. This will reduce the total number of writes required to cause spy slowdown and as a result, improve the bit rate. 

\begin{figure}[t]
    \centering
    \includegraphics[width=0.4\textwidth]{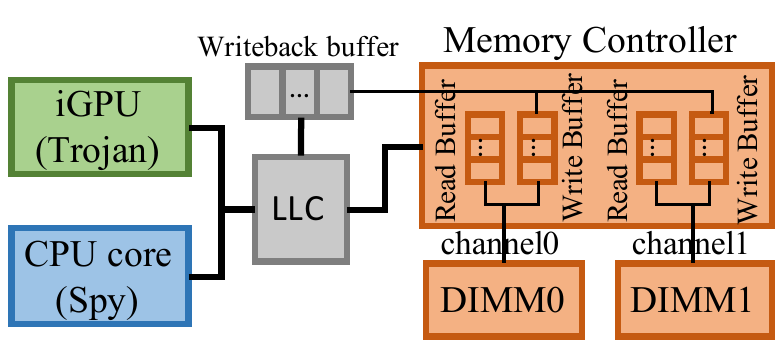}
    \caption{Attack Variant 2: Targeting Single MC Channel.}
    \label{fig:channel_targeted_attack}
\end{figure}

Similar to the experiments we have performed in Section~\ref{subsec:var1_design}, we inferred the parameters required for this attack approach. The trojan buffer should be accessed at a stride of four rather than eight. This is because half of iGPU kernel writes will access targeted channel, and we need to ensure eviction of dirty cache lines. Note that the total number of iGPU kernel writes in this attack is similar to the previous attack. The difference is that iGPU kernel traffic is directed to one pre-agreed channel.
In this attack, the iGPU kernel write requests are preceded with a buffer index read. This is the buffer index that iGPU kernel will be writing to next and which is targeting the same channel as spy reads. Because of this, the frequency of writes is lower than the previous attack and bit '0' iteration factor of two has better attack performance. Local threads of 128 and global threads of 256 achieved a good performance in terms of error and bit rates similar to previous attack.

The bit rate in this attack is higher because bit '0' iteration factor is smaller. Additionally, frequency of write requests from iGPU kernel is lower and as result some of spy read requests will suffer lower latency than the latency we show in Figure~\ref{fig:latency_per_access}.

\section{Covert Channel Attacks Evaluation}
\label{sec:eval}
In this section, we explain the attack evaluation setup and evaluate both attack variants in terms of bit and error rates. 

\subsection{Experimental Setup}
We target an Intel-based SoC with an integrated GPU. Most desktop and mobile Intel processors include integrated GPU for the acceleration of graphics and compute workloads\cite{gen9,gen11, tap_georgiatech_hpca}. All our experiments are on a Cometlake i7-10700K processor, which features an integrated Intel UHD Graphics 630 (based on gen9.5 architecture with 1 slice of 3 subslices, and 8 EUs per subslice). We use OpenCL version 2.1, running Ubuntu version 20.04.1 (which uses Linux Kernel version 5.13). The size of LLC and main memory are 16MB and 32GB, respectively. The SoC has a single memory controller with dual channels. Table~\ref{tab:exp_setup} summarizes the attacks' experimental setup.

\begin{table}[t]
\centering
\begin{tabular}{||m{0.4\columnwidth} || m{0.5\columnwidth}||}
    \hline
     CPU &  8 cores, Cometlake i7-10700K @ 3.80GHz\\
     \hline
     L1 \& L2 Caches & L1-I\$,L2-D\$:32KiB (8-way)\newline L2\$:256KiB (4-way) \\
     \hline
     LLC & 16MiB (16-way)\\
     \hline
     iGPU &  Intel UHD Graphics 630 (gen9.5) @ 1200MHz\\
     \hline    
     Mem. Controller (MC) & one MC, dual channel\\
     \hline
     DRAM & 32GB\\
     \hline
\end{tabular}
\caption{Attacks environment setup.}
     \label{tab:exp_setup}
     \end{table}

\subsection{Evaluation of Attack Variant 1 }
\begin{figure}[t]
\hspace{-5mm}
\subfloat[\label{fig:attack_var1}]{\includegraphics[width=0.55\columnwidth]{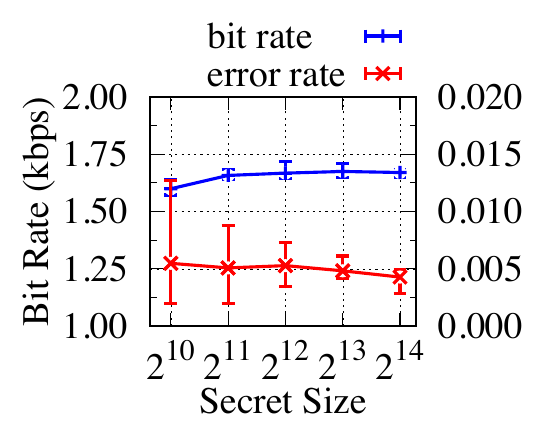}}
\subfloat[\label{fig:attack_var2}]{\includegraphics[width=0.55\columnwidth]{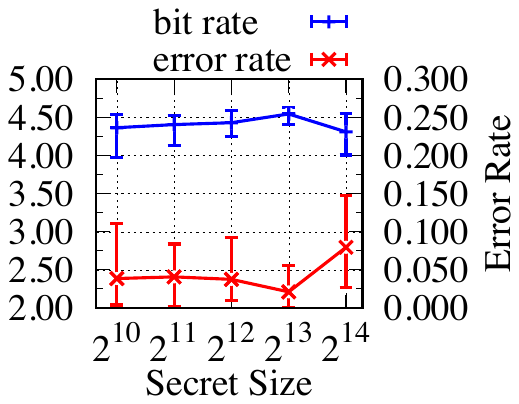}}
\caption{Bit and error rates for (a) attack variant 1 (b) and attack variant 2.}
\label{fig:bit_err_rates}
\vspace{-2mm}
\end{figure}

This attack depends on allocating a large buffer that is accessed by the iGPU kernel to evict dirty cache lines. Figure~\ref{fig:attack_var1} depicts both bit and error rates. The parameters are based on our analysis in section~\ref{subsec:var1_design}. The attack can achieve an average bit rate of 1.65 kbps with an average error rate of 0.49\%. This is achieved when the iGPU kernel buffer is accessed at a stride of eight cache lines. If a stride of 16 cache lines is selected, an average bit rate of 2.43 kbps is possible but with an average error rate of 29.87\%.



\subsection {Evaluation of Attack Variant 2}
This attack requires spy and trojan to pre-agree on a specific MC channel to access. This requires channel bit addressing knowledge to know which bit in physical address determine which channel to access. We evaluated this attack variant based on the parameters we have inferred in section~\ref{subsec:attack_var2}. Figure~\ref{fig:attack_var2} depicts bit and error rates for different secret sizes. This attack variant achieved an average bit rate of 4.41 kbps and average error rate of 4.32\%. Attack variant 2 achieved bit rate $\times2.67$ larger than attack variant 1, but average error rate is $\times8.8$ larger than attack variant 1.

\section{Discussion}
\label{sec:discuss}
\noindent\textbf{Characterizing Bit and Error Rates.} Our covert channel attack variants depend on memory writes. From Figure~\ref{fig:latency_per_access}, we can notice the high latency of spy reads due to iGPU kernel writes. Such high latency contributes to lowering the bit rate depending on the frequency of iGPU kernel writes and the number of affected spy read requests. It can also be the reason behind the low error rate because the difference between bit '0' and bit '1' latency is quite high.

\noindent\textbf{Attack Generalization.} We discuss the possibility of generalizing our covert channel attack to other integrated accelerators. Our attack mainly depends on frequently filling the write buffer in the MC to stall spy (CPU process) memory read requests. If the system is adopting management policies for write buffers such as \textit{drain\_when\_full} or if it delays memory read requests upon frequent filling of memory write buffer, then our attack is applicable to such systems. 
Not all integrated accelerators share the LLC with CPU cores like Intel iGPUs. Current AMD APUs do not share LLC with iGPU~\cite{cezanne}. In such a system, because the size of caches in iGPU is much smaller than LLC, we expect better attack performance. This is because it is possible to frequently fill the write buffer using a smaller buffer and random evictions. Moreover, if deterministic evictions are chosen, this will speed up the process of finding an eviction set.

\noindent\textbf{Degree of iGPU Parallelism.}
From Figure~\ref{fig:glob_loc_threads} we noticed that increasing the local work-group size lowered the error rate but increased the spy execution time. Also, an increasing number of local work-groups beyond two groups increases the error rate but lowers the spy execution time. In our targeted system, the maximum size of a local workgroup is 256 work-items. From such results, we observe that increasing the number of work-groups does not necessarily improve the attack performance. This is due to the performance overhead associated with creating and scheduling these work-groups. Furthermore, a larger number of local work-groups means a lower number of writes per thread, if all global threads are used. This will cause the time required for scheduling to dominate, and thus reducing the frequency of writes which will cause the error rate to increase. Consequently, there has to be a consideration of the number of available resources such as the number of EU (Compute Units) and created threads.


\section{Possible Mitigations}
\label{sec:mitigations}
In this section, we discuss different possible mitigations to thwart the attacks or to make it challenging for attackers to exploit memory write requests.\newline
\noindent{\textbf{Prioritizing Memory Read Requests.}} One of the possible solutions for the latency observed by the CPU process due to iGPU kernel memory write requests is deploying a policy other than \textit{drain\_when\_full}. Deploying a management policy that gives priority to memory read requests whenever they are inserted in the read buffer will thwart our attack. Such an approach may prevent iGPU kernel or CPU processes from issuing write requests if there are read requests in the read buffer and the write and writeback buffers are full. This will deteriorate the performance of CPU processes or iGPU kernel if they are issuing a lot of memory write requests. Also, latency overhead will appear due to \textit{read\_after\_write} and \textit{write\_after\_read} especially if there is frequent alteration between serving memory read and write requests.

\noindent{\textbf{Directing CPU Process's Memory Reads to Ideal DRAM Banks.}} Chatterjee et al.~\cite{staged_reads} proposed Staged Reads to serve memory read requests while memory write requests are being served. This happens only if memory reads are accessing banks not used by memory writes (i.e. ideal banks). Such an approach will not completely thwart our attack, but it will make it challenging for attackers to utilize the effect of \textit{drain\_when\_full} management policy. If Staged Reads \cite{staged_reads} technique is deployed in the system, spy and trojan should pre-agree on accessing same banks for spy read requests to be stalled.

\noindent{\textbf{Channel Partitioning.}} It is possible to thwart our covert channel attack using channel partitioning in multi-channel MCs. This is because each MC channel has its own pair of read and write buffers. One of the MC channels can be dedicated to iGPU during its kernel execution. If the CPU process needs to read from memory associated with this channel, it has to stall until the iGPU kernel finishes execution. Obviously, this approach has a performance impact on CPU processes, especially in the case of kernels with long execution times.

\section{Related Work}
\label{sec:related_work}
We discuss related research to our proposed attacks. Such research either exploited iGPUs, memory controller resources, or write requests. 

\noindent\textbf{Attacks Targeting Memory Controller's and DRAM Resources.}
DRAMA~\cite{drama_usenix} targeted row buffer conflicts and exploit it for covert and side channel attacks. Their approach required attack parties to agree on a set of channel, rank, and bank. Similar to DRAMA, Xiao et al.~\cite{xiao_ohio} reversed engineering bank bit addressing, but their attack is for double-sided row hammer in virtualized environments.
Wang et al.~\cite{cornell_defense_hpca} demonstrated the possibility of contention-based covert and side channel attacks to motivate their defense scheme. Their side channel attack targeted RSA decryption algorithm by causing a cache miss when a modulo operation is executed. Their work neither developed end-to-end attacks nor reverse-engineered the source of contention. Unlike these attacks, our attack mainly exploits the management policy of the write buffer in MC.

\noindent\textbf{iGPU as an Attack Vector.} Some existing work exploited integrated accelerators to infer secret data. They are also used to speed up the process of known attacks such as finding eviction sets or DRAM bit flipping. Leaky Buddies~\cite{leaky_buddies} used iGPU in their covert channel attacks which targeted LLC and Ring Interconnect. 
Our attack does not target LLC or ring-interconnect. Grand Pawning Unit~\cite{Glitch-2018} used integrated accelerators (GPUs) for accelerating bit flipping used in attacks such as rowhammer and breaking ASLR. 

\noindent\textbf{Attacks Targeting Cache/Memory Writes.} Most of the existing side and covert channel attacks targeted cache/memory reads rather than writes (temporal writes). CPU processes can not time cache/memory writes beyond updating associated cache lines; therefore they can not reveal much of information about the program's behavior, especially at the MC level. Also, approaches such as write-no-allocate and write-back caches can exacerbate this case. Thoma et al.~\cite{raid_cpu_write} proposed Write-After-Write side channel attack where they noticed that writing to an address in cache can make subsequent writes slower to addresses colliding with the previously written address.

\section{Conclusion}
\label{sec:conc}
In this paper, we characterized the slowdown observed in the CPU process due to iGPU kernel memory writes. We showed that it is possible to cause a very large slowdown to CPU process memory reads by just utilizing 1/3 of iGPU available resources to issue memory write requests. We confirmed that this slowdown happens due to the management policy of write buffer and not any other shared resources (i.e. LLC, ring-interconnect, or DRAM banks). We proposed two covert channel attack variants which exploit this phenomenon to leak secret information. The first variant is oblivious of the used MC channel, while the second variant is targeting a specific MC channel to improve the bit rate 

We believe our attacks demonstrate a critical leakage vector across the components in modern heterogeneous systems that guide future research into designs that are not only high-performance but also secure. 

\bibliographystyle{plain}
\bibliography{main}

\end{document}